\documentclass[a4paper,preprint,groupedaddress,showpacs,nofootinbib]{revtex4}

\usepackage{amssymb,amsmath,graphicx}
\usepackage[dvipsnames]{color}

\begin{document}

\title[Is the Hubble Constant Scale-dependent?]%
      {Is the Hubble Constant Scale-dependent?}

\author{Yurii V. Dumin}

\email[]{dumin@yahoo.com, dumin@sai.msu.ru}

\affiliation{
Sternberg Astronomical Institute (GAISh), Lomonosov Moscow State University,\\
Universitetskii prosp.\ 13, Moscow, 119992 Russia}

\affiliation{
Space Research Institute (IKI), Russian Academy of Sciences,\\
Profsoyuznaya str.\ 84/32, Moscow, 117997 Russia}

\begin{abstract}
An exact determination of the Hubble constant remains one of key
problems in cosmology for almost a century.
However, its modern values derived by various methods still disagree from
each other by almost~10{\%}; the greater values being obtained
by measurements at the relatively small distances (\textit{e.g.}, by Cepheid
stars as the standard candles), while the smaller values being characteristic
of the methods associated with huge spatial scales (\textit{e.g.},
the analysis of the cosmic microwave background fluctuations).
A reasonable way to resolve this puzzle is to assume that the Hubble
constant is inherently scale-dependent.
This idea seems to be particularly attractive in light of
the latest observational results on the early-type galaxies, where
the dark-matter haloes are almost absent.
Therefore, an average contribution of the irregularly-distributed dark matter
to the rate of the cosmological expansion should be substantially different
at various spatial scales.
As follows from the rough estimates, the corresponding variation of the Hubble
constant can be~10{\%} and even more, which well explains
the spread in its values obtained by the various methods.
\end{abstract}

%
% 98.80.Es Observational cosmology (including Hubble constant, distance scale,
%          cosmological constant, early Universe, etc)
% 95.35.+d Dark matter (stellar, interstellar, galactic, and cosmological)
% 95.36.+x Dark energy
% 98.62.Gq Galactic halos

\maketitle

As is known, measurement of the Hubble constant~$ H_0 $ represents
a long-standing problem of cosmology, whose history lasts for almost a century.
The resulting values, obtained in this period, varied by an order of magnitude,
50 to 500~km/s/Mpc~\cite{Ryden_2017}.
Although the situation improved in the recent decades, some discrepancies
persist till now.
The most notable of them is that the value of~$ H_0 $ derived from the distance
scale based on Cepheids is, on the average, $ 73.24{\pm}1.74 $~km/s/Mpc and for
some calibration can even be as large as~$ 76.18{\pm}2.37 $~\cite{Riess_2016}.
On the other hand, the analysis based on measurements of the cosmic microwave
background (CMB) by Planck satellite under assumption of the $\Lambda$CDM
cosmological model leads to the values $ H_0 = 66.88{\pm}0.91 $ to
$ 67.31{\pm}0.96 $~km/s/Mpc, depending on the method of data
processing~\cite{Aghanim_2016}.
In other words, these numbers are about 10{\%} less than in the first case.%
\footnote{
This should not be confused with the fact that the Hubble parameter
is a continuously decreasing function of cosmological time.
So, its value is larger for the remote galaxies, since we always look
into the past.}

The above-mentioned discrepancy between the direct (by Cepheids) and indirect
(by CMB) measurements of~$ H_0 $ was clearly recognized in the recent years;
and it is commonly attributed now either to the systematic errors (such as
degeneracy between different quantities in the analysis of CMB) or to
the uncertainty in the fitting parameters (\textit{e.g.}, the number and masses
of neutrinos, \textit{etc.}).
An especially popular explanation became a modification of the parameter~$ w $,
appearing in the dark-energy equation of state, $ p = w \varepsilon $ (see,
for example, paper~\cite{Huang_2016} and references therein); though
the resulting values $ w < -1 $ look quite unrealistic and suspicious from
the physical point of view.%
\footnote{
For example, the values of~$ w $ somewhat greater than~$ -1 $
(\textit{i.e.}, $ |w| < 1 $) could be easily interpreted as a result of the
small-scale spatial irregularities in the equation of state of the scalar
field representing the dark energy~\cite{Linde_1984}, but such an effect
cannot lead to $ w < -1 $.}

However, from our point of view, the spread in values of~$ H_0 $ can have
a much more straightforward astrophysical explanation: this quantity
should be inherently scale-dependent.
Really, according to the Friedmann equation, the Hubble constant
depends on the energy density in the Universe as~\cite{Olive_2016}:
\begin{eqnarray}
H_0 & = & \sqrt{ \frac{ 8 {\pi} G }{ 3 }} \,
    \sqrt{ \, {\rho}_{\rm DE} + {\langle} {\rho}_{\rm DM} {\rangle}
    + \dots } \; =
\nonumber \\[1ex]
&&
\sqrt{ \frac{ 8 {\pi} G {\rho}_{\rm DE} }{ 3 }} \: \sqrt{ \, 1 + 
    \frac{ {\langle} {\rho}_{\rm DM} {\rangle} }{ {\rho}_{\rm DE} }
    + \dots } \:\: ,
\label{eq:Hubble_param}
\end{eqnarray}
where $ G $~is the gravitational constant,
$ {\rho}_{\rm DE} $~is the density of the dark energy,
which is assumed to be distributed perfectly uniform in space,
$ {\langle} {\rho}_{\rm DM} {\rangle} $~is the average density of
the dark matter (whose value depends on the scale of averaging),
and dots denote the contribution from ordinary forms of matter,
which is not greater than~5{\%} (and, consequently, its contribution to
the value of Hubble constant will be about~2.5{\%}).

Since both the physical origin and spatial distribution of the dark matter are
actually unknown by now~\cite{Liu_2017,Buchmueller_2017}, it can be naturally
assumed that its contribution to formula~(\ref{eq:Hubble_param}) substantially
depends on the spatial scales under consideration.
Particularly, according to the recent observational
findings~\cite{Swinbank_2017,Genzel_2017}, the dark matter is almost absent
in the vicinity of early-type galaxies, located at large
redshifts~$ z \approx 0.6{-}2.6 $.
Therefore, averaging over the larger spatial scales should result in
the smaller values of~$ {\langle} {\rho}_{\rm DM} {\rangle} $.

As follows from expression~(\ref{eq:Hubble_param}), the corresponding
variance of the Hubble parameter,
$ {\delta}H_0 = H_0^{\rm (max)} - H_0^{\rm (min)} $,
can be as large as
\begin{equation}
{\delta}H_0 / H_0^{\rm (max)} \approx \, \frac{1}{2} \,
  {\big( {\rho}_{\rm DM}^{\rm (max)} / {\rho}_{\rm DE} \big)} ,
\label{eq:variance}
\end{equation}
where $ H_0^{\rm (min)} $ formally corresponds to
$ {\langle} {\rho}_{\rm DM} {\rangle} = 0 $,
and $ H_0^{\rm (max)} $ to
$ {\langle}{\rho}_{\rm DM}{\rangle} = {\rho}_{\rm DM}^{\rm (max)} $.

Therefore, taking for estimate
$ {\rho}_{\rm DM}^{\rm (max)} / {\rho}_{\rm DE} \approx 3/7 $~\cite{Lahav_2016},
we find that the relative variance $ {\delta}H_0 / H_0^{\rm (max)} $ can reach
approximately~20{\%}.
In fact, the realistic value should be somewhat less, because
the above estimate was obtained under the simplifying assumption that
$ {\langle} {\rho}_{\rm DM} {\rangle} \to 0 $ at very large scales.
Anyway, the systematic 10{\%}~discrepancy in the values of~$ H_0 $ derived by
the various methods is not surprising: a ``direct'' determination of the Hubble
constant from the extragalactic distance scale based on the Cepheid
variable stars refers to the relatively local part of the Universe; while
the ``indirect'' analysis based on the CMB fluctuations deals with much larger
scales.
As was already mentioned in the above-cited work by Genzel,
et al.~\cite{Genzel_2017}, at such scales the dark matter should play a smaller
part than in the local Universe.

It is important to emphasize that, since both the dark and luminous matter
possess the same dust-like equation of state ($ w \approx 0 $), their temporal
evolution (in the cosmological sense) should be the same, \textit{i.e.} the
ratio of their densities should be constant. So, the deficit of
dark matter in the vicinity of high-redshift galaxies cannot be
explained merely by the fact that they are observed at earlier times.

In summary, we believe that the well recognized discrepancy between
the various determinations of~$ H_0 $ could be more naturally explained by
the irregularities of matter distribution, not taken into account explicitly
in the standard Friedmann equation, rather than by modifications
of the equation of state for the dark energy or other exotic assumptions
in the framework of ``uniform'' cosmological equations.
(In fact, this issue is closely related to the general problem of
``excessive extrapolations in cosmology'', which was pictorially outlined
in the recent paper~\cite{Krizek_2016}.)


\begin{thebibliography}{99}

\bibitem{Ryden_2017}
B.~Ryden,
``A Constant Conflict'',
\textit{Nature Physics} {\bf 13} 314 (2017).

\bibitem{Riess_2016}
A.G.~Riess, L.M.~Macri, S.L.~Hoffmann, D.~Scolnic, S.~Casertano, et al.,
``A 2.4\% Determination of the Local Value of the Hubble Constant'',
\textit{Astrophys.\ J.} {\bf 826} 56 (2016).

\bibitem{Aghanim_2016}
N.~Aghanim, M.~Ashdown, J.~Aumont, C.~Baccigalupi, M.~Ballardini, et al.\
(Planck Collaboration),
``Planck Intermediate Results: XLVI. Reduction of Large-Scale Systematic
Effects in HFI Polarization Maps and Estimation of the Reionization
Optical Depth'',
\textit{Astron.\ Astrophys.} {\bf 596} A107 (2016).

\bibitem{Huang_2016}
Q.-G.~Huang and K.~Wang,
``How the Dark Energy Can Reconcile Planck with Local Determination of
the Hubble Constant'',
\textit{Eur.\ Phys.\ J.\ C}
{\bf 76} 506 (2016).

\bibitem{Linde_1984}
A.D.~Linde,
``The Inflationary Universe'',
\textit{Rep.\ Prog.\ Phys.}
{\bf 47} 925 (1984).

\bibitem{Olive_2016}
K.A.~Olive and J.A.~Peacock,
``Big-Bang Cosmology'',
in: C.~Patrignani, et al.\ (Particle Data Group),
\textit{Review of Particle Physics},
\textit{Chinese Physics C} {\bf 40} 100001 (2016), p.355.

\bibitem{Liu_2017}
J.~Liu, X.~Chen, and X.~Ji,
``Current Status of Direct Dark Matter Detection Experiments'',
\textit{Nature Physics} {\bf 13} 212 (2017).

\bibitem{Buchmueller_2017}
O.~Buchmueller, C.~Doglioni, and L.-T.~Wang,
``Search for Dark Matter at Colliders'',
\textit{Nature Physics} {\bf 13} 217 (2017).

\bibitem{Swinbank_2017}
M.~Swinbank,
``Distant Galaxies Lack Dark Matter'',
\textit{Nature} {\bf 543} 318 (2017).

\bibitem{Genzel_2017}
R.~Genzel, N.M.~F{\"o}rster Schreiber, H.~{\"U}bler, P.~Lang, T.~Naab, et al.,
``Strongly Baryon-Dominated Disk Galaxies at the Peak of Galaxy Formation
Ten Billion Years Ago'',
\textit{Nature} {\bf 543} 397 (2017).

\bibitem{Lahav_2016}
O.~Lahav and A.R.~Liddle,
``The Cosmological Parameters'',
in: C.~Patrignani, et al.\ (Particle Data Group),
\textit{\it Review of Particle Physics},
\textit{Chinese Physics C} {\bf 40} 100001 (2016), p.386.

\bibitem{Krizek_2016}
M.~K\v{r}\'{i}\v{z}ek and L.~Somer,
``Excessive Extrapolations in Cosmology'',
\textit{Grav.\ Cosmol.}
{\bf 22} 270 (2016).

\end{thebibliography}
\end{document}